\documentstyle[preprint,aps,eqsecnum]{revtex}
\begin{document}
\newcommand {\be}{\begin{equation}}
\newcommand {\ee}{\end{equation}}
\newcommand {\la}{\label}
\newcommand {\r}{\ref}
\newcommand {\bea}{\begin{eqnarray}}
\newcommand {\eea}{\end{eqnarray}}

\title
{ Raman Scattering and Anomalous Current Algebra:
Observation of Chiral
Bound State
in Mott Insulators}

\author{D.V.Khveshchenko $^{a,c}$ and P.B.Wiegmann $^{b,c}$\footnote
{e-mail: {\it wiegmann@control.uchicago.edu}}}
\address
{$^a$ Department of Physics, Princeton University,\\
Princeton, NJ 08544\\
$b$ James Franck Institute and Enrico Fermi Institute and \\Department of
Physics of the  University of Chicago,\\ 5640, S.Ellis Ave., Chicago, Il
60637\\
$c$ Landau Institute for Theoretical Physics,
2, Kosygina st., Moscow 117940, Russia}
\maketitle

\begin{abstract}
\noindent

Recent experiments on inelastic
light scattering
in a number of
insulating cuprates [1]
revealed a new excitation
appearing in the case of crossed polarizations
just below the optical absorption threshold.
 This observation suggests that
there exists a local exciton-like state with an
odd parity with respect to a spatial reflection. We present the
theory of high energy large shift  Raman scattering in Mott
insulators and interpret the
experiment [1]  as an evidence of
 a chiral bound state of a hole and a doubly occupied site with a
topological magnetic excitation. A formation of these composites
is a crucial feature
of various topological mechanisms of superconductivity.
 We show that inelastic light scattering  provides an instrument for  direct
 measurements of a local chirality and
anomalous terms in the electronic current algebra.
\end{abstract}
\pagebreak

1. {\it Introduction}

Recent Raman studies of a number of
insulating cuprates revealed a resonant feature
just below the optical  absorption peak [1].
Remarkably, the most robust and
prominent feature  has been
observed in crossed polarizations, i.e. in a
scattering
 geometry corresponding to a pseudovector symmetry
(so-called $A_2$ geometry). The
present comment is motivated by these observations which
suggest that

(i) within the Mott-Hubbard gap there is an exciton-like bound state,

(ii) this
  bound state has an odd symmetry under reflection of a two dimensional
square lattice.

Namely, we argue that the results of the experiments [1]
can be interpreted as an evidence of a chiral character of charged excitations
in cuprates.

 Raman data have already yielded an important information about magnetic
and phonon  properties of insulating cuprates
[2-5]. Furthermore, new experiments on {\it high energy, large
shift } Raman scattering [1]
 provide a powerful tool for a study of charge excitations in
Mott insulators.
These data  might also
give the most valuable information for mechanisms of
superconductivity in doped materials.
In this paper we show that this technique presents a
direct way to investigate
the electronic current algebra and a chirality of excitations
which are
crucial features of various topological mechanisms of superconductivity [6,7].

2. {\it Raman scattering cross section}

A process of inelastic Raman
scattering is an absorption of an incident photon with a frequency
$\omega_i$,
a wave vector $\vec k_i$, and a
polarization $\vec e_i$   and a simultaneous
emission  of a scattered photon ($\omega_f,\vec k_f,\vec e_f$).
In the case of insulator the Raman scattering cross section is given by the
Kramers-Heisenberg formula [8]:
\begin{equation}
R(\Omega)={{\omega^3_f \omega_i}\over{2\pi h^2c^4}}\sum_{i,f}e^{-\beta
E_i}|M_{if}|^2\delta(E_f-E_i+h\Omega)
\label{R1}
\end{equation}
where
\begin{equation}
M_{if}=\sum_n {<i|j_i|n><n|j_f|f>\over{E_n-E_i-\omega_i}}+
{<i|j_i|n><n|j_f|f>\over{E_n-E_i-\omega_f}}
\label{M1}
\end{equation}
is the scattering tensor. Here the sum goes over all intermediate states
$ |n>$ with energies $E_n$   and $j_{i,f}=\vec j(k_{i,f}){\vec e}_{i,f}$ is
an
electromagnetic current operator in the direction of a polarization of
incident
(scattered) photon. As a function of energy
($\Omega=\omega_i-\omega_f $)
 and momentum ($\vec q=\vec k_f-\vec k_i$) transfer the
scattering rate can be written as a
correlation function
of a time dependent scattering tensor
\begin{eqnarray}
&&M=e_i^{\mu} e_f^{\nu}M_{\mu \nu}({\vec r},\tau),\nonumber\\
&&M_{\mu \nu}({\vec r},\tau)=\int
e^{-i\Omega \tau-i\vec q\vec r}
M^{\mu \nu}(\vec q,\Omega)d\vec q d\Omega=\nonumber\\
&&=i
\int d{\vec r}^{\prime}\int^{\infty}_{0}d\tau^{\prime}
e^{-i\omega_{i}\tau^{\prime}+i\vec k_i{\vec r}^{\prime}} [j_{\mu}({\vec
r}+{\vec r}^{\prime},\tau+\tau^{\prime}),j_{\nu}({\vec r},\tau)]
\la{M1}
\end{eqnarray}
Omitting the prefactor in the Eq.(\ref{R1}) we obtain at zero temperature
\be
R(\vec q, \Omega) \sim Im<i| M^{\dag}(\vec q,\Omega)M(\vec q,\Omega)|i>
\la{R2}
\ee
Since photon wavelengths are always
 much larger than an interatomic separation
one may neglect a spatial dispersion of the scattering tensor
putting $\vec q=0$ in Eqs.(3,4).

In general,
scattering rates appear to be strongly dependent on photon polarizations.
The scattering tensor (3) can be
decomposed into four one dimensional
irreducible representations of the square lattice point group $D_{4h}$ [4]:
\bea
M_{A_1}=M_{xx}+M_{yy},\nonumber\\
M_{B_1}=M_{xx}-M_{yy},\nonumber\\
M_{A_2}=M_{xy}-M_{yx},\nonumber\\
M_{B_2}=M_{xy}+M_{yx}
\eea
Below we shall draw the main
attention to the $A_2$  scattering amplitude which is odd under
reflection on the square lattice and gives the strongest signal in the
experiments [1].

3.{\it High energy Raman scattering  and the current algebra}

In experiments [1] the frequency of the incident
light was $\omega_i=3.4-3.8 ev$ and the
$A_2$ resonance was found at Raman shift $\Omega\approx 1.5ev$   -
which is about $0.15 - 0.20ev$ below the optical absorption peak.
 To elucidate the mechanism of the $A_2$ scattering
 we exaggerate the conditions of the experiment, assuming that energies of
incident and scattered light are
much larger than the widths of relevant electronic bands,
so one can
neglect the dispersion of $E_n$ in denominators of the Eq.(2).

In this limit the  time
separation between current operators in the Eq.(3) is very small
 and in the leading approximation the scattering
tensor is given by an
equal-time current-current commutator. Apparently, a nontrivial
scattering occurs only in the $A_2$ geometry:
\be
M\approx M_{A_2}={1\over{\omega_i}}[j_x(\tau),j_y(\tau)]
\la{M4}
\ee
where $\vec j(\tau)=\int d\vec r \vec j(\vec r,\tau)$
 is the spatial average of the current operator.

Thus, we conclude that at large laser energy the
Raman  scattering in the insulator
is only due to a nonzero value of the  equal time current-current commutator
\begin{eqnarray}
\la{R4}
R(\Omega)\approx R_{A_{2}}(\Omega)=
{\omega_{f}^{3}\over{\omega_i {2\pi h^2 c^4}}}
\int d\tau e^{i\Omega \tau}
<0|[j_{x}(\tau),j_{y}(\tau)][j_{x}(0),j_{y}(0)]|0>
\end{eqnarray}
 Scattering rates in even geometries $A_1$ and $B_2$ appear in the
second order in $\omega^{-1}_i$  while the amplitude
 of the $B_1$ scattering is even smaller:
\bea
M_{A_1}={1\over{i\omega_i^2}}
([{d\over{d\tau}}j_x(\tau),j_x(\tau)]+[{d\over{d\tau}}j_y(\tau),j_y(\tau)]),
\nonumber\\
M_{B_2}={1\over{i\omega_i^2}}
([{d\over{d\tau}}j_x(\tau),j_y(\tau)]+[{d\over{d\tau}}
j_y(\tau),j_x(\tau)]),\nonumber\\
M_{B_1}={1\over{\omega_i^3}}
([{d^2\over{d\tau^2}}j_x(\tau),j_x(\tau)]-[{d^2\over{d\tau^2}}j_y(\tau),
j_y(\tau)])
\la{M4}
\eea
Below we shall
ignore any intraatomic structures and apply the simplest two band
Hubbard Hamiltonian  as a model of the Mott insulator. Then
${d\over{d\tau}}\vec j=-i[H,\vec j]$ can be easily calculated. Introducing
one fermion translation
operators $T_\mu(\vec r)=\sum_{\sigma=\uparrow,\downarrow}
 c^{+}_\sigma (\vec r+\hat \mu) c_\sigma (\vec r)$
we express the current as
 $j_\mu (\vec r)={{et}\over {2ihc}}(T_\mu(\vec r)-T_{-\mu}(\vec r))$. Then
keeping terms up
to the third order in $\omega_i^{-1}$ we obtain
\begin{eqnarray}
&&M_{A_2}={1\over{\omega_i}}[j_x(\tau),j_y(\tau)](1+{{U^2}\over
{\omega_{i}^2}})+
{{t^2}\over {\omega_{i}^3}}\bigl(j_x(\tau)(\sum_{\vec r,\mu}
T_\mu (\vec r))^2
j_y(\tau)-j_y(\tau)(\sum_{\vec r,\mu}T_\mu (\vec r))^2 j_x(\tau)\bigr),
\nonumber\\
&&M_{A_1}=-{2U\over{\omega_i^2}}({j^2}_x(\tau)+{j^2}_y(\tau)),\nonumber\\
&&M_{B_2}=-{2U\over{\omega_i^2}}\{j_x(\tau),j_y(\tau)\}
\la{M4}
\end{eqnarray}
where $U$ is the energy of the onsite Coulomb repulsion and $t$ is the hopping
amplitude (We kept the last term in the formula for $M_{A_2}$ because it is the
only one which contributes to the
 rate of the $A_2$ quasielastic  scattering ($\Omega\rightarrow 0$).
The latter has been
considered by Shraiman and Shastry [4] under assumption
 $|U-\omega_i|<< \min{(U,\omega_i)})$.

In
what follows  we shall omit the absolute scale of the scattering rates
given by the factor
${\omega_{f}^{3}\over {2\pi\omega_i h^2 c^4}}
({e\over{2hc}})^4$ and set
${e\over{2hc}}=1$.

4. {\it Chirality operator}

A charge excitation in the Mott insulator involves a distortion of the magnetic
state. As a result, a phase of the excitation wavefunction
 depends not only on
 its current
location but also on a path, which the hole passed to arrive at
the site. Therefore a hole inserted to the Mott insulator
may acquire a phase while moving along the closed
path. This property is the quantum holonomy
which is determined by a nonzero spin chirality of the
insulating
magnetic state [9,10]. We define it quantitatively  as a measure
of noncommutativity of one electron translation operators $T_\mu$
as applied to the insulating ground state
$$ T_xT_y=e^{i {\hat{\Phi}}}T_yT_x = e^{{i\over 2}
{\hat{\Phi}}}T_{x+y}$$
The chiral  operator $\exp{i {\hat{\Phi}}}$
is defined as  a product of
translation operators over a closed lattice contour $C$. On the other hand
 it can be
expressed in terms of spin
operators by means of the relation
\be \prod_C T_{\mu}(\vec r)=tr\prod_C(1+2\vec\sigma \vec S(\vec r))\la{TT}\ee
We  show that
high energy Raman scattering is an instrument to measure matrix elements of
the chirality operator.

5. {\it Integrated intensity and  elastic scattering}

Some information about spin chiralities can be already
obtained from
the
integrated Raman intensity $R^{int}=\int_0^\infty d\Omega R(\Omega)$
and from the elastic scattering
rate $R^{el}(\Omega)=R(\Omega\rightarrow 0)$.
In the leading order of $\omega_i^{-1}$
the integrated intensity is given by the formula
\be
\la{integrate}
R^{int}\approx R^{int}_{A_{2}}\sim <0||[j_{x},j_{y}]|^2|0>
\ee
Since in (11) all four current operators appear at the same
moment of time they
have to
 be located in adjacent points on a plaquette to
form a
four step closed contour. Then it is easy to see that
the  integrated intensity measures
fluctuations of a local chirality. After a simple algebra we
obtain
\be
\la{int}
R^{int}\sim <0|(\sin{{1\over 2} {\hat{\Phi}} })^2|0>
\ee
It can be also expressed in terms of spin operators
 \be
\la{integrate1}
R^{int}\sim 4\sum_{P} <0|1-3/2 (\vec S_a\vec\cdot S_{a^{\prime}})
+2(\vec S_a\vec S_b)(\vec S_a\vec\cdot S_{b^{\prime}})
-(\vec S_a\vec S_{a^{\prime}})(\vec S_b\vec\cdot S_{b^{\prime}})|0>
 \ee
where $a,a^{\prime}$ and $b,b^{\prime}$ are the nearest neighbouring sites
on $A$ and $B$ sublattices.
The integrated intensity in other scattering geometries is smaller and
does not contain any information about chirality :
 $ R^{int}_{A_{1}}\sim
R^{int}_{B_{2}}\sim (U/ \omega_i)^2$.

 The amplitude of an elastic
scattering is a more informative object.
In the leading order in $\omega_i^{-1}$ it
measures an averaged difference between holonomies  associated with
two oppositely oriented elementary
closed contours which is equal
to the average chirality of the ground state. From
the Eq.(9) we obtain
\be
\la{elastic1}
R^{el}(\Omega)\sim |<0|M_{A_2}|0>|^2\delta(\Omega)
\approx
({t\over{\omega_i}})^4|\sum_{P}<0|T_xT_yT_{-x}T_{-y}-T_{-x}T_yT_{x}T_{-y}|0>|^2
\delta(\Omega) \ee
where the sum goes over all plaquettes.
The matrix element staying in the Eq.(14)
yields the total solid angle formed by
all spins in the ground state
\be
\la{elastic2}
R^{el}(\Omega)\sim
({t\over{\omega_i}})^4{\bigl(\sum_{P}<0|(\vec S_1\times \vec S_2)\cdot \vec
S_3|0>\bigr)}^2 \delta(\Omega) \ee
where the triple $1,2,3$ denotes any three nearest
neighbouring sites on the plaquette $P$.

 In other geometries an elastic scattering appears only
in the next order in $\omega_i^{-1}$ and
doesn't  contain holonomy.  The intensity of the
elastic $A_1$ scattering  has
 the same order of magnitude
as the corresponding
integrated  intensity $R^{el}_{A_{1}}\sim R^{int}_{A_{1}}$
while the
$B_2$ elastic rate is much smaller then the integrated one
$R^{el}_{B_{2}}\sim  {{t^4}\over{U^2\omega_i}}R^{int}_{B_{2}}$.

An observation of a separate peak in
the $A_2$ elastic scattering would mean a spontaneous parity breaking in the
ground state
[9-11]. Despite of
a small amplitude, a peak in the elastic
scattering (if any), could be detected experimentally. However it has never
been observed. Although an  additional experimental study is desirable, it
seems
most likely  that  the ground state of the half-filled Mott
insulator is parity even.

6. {\it Resonant scattering}

The most interesting information can be extracted
from the resonant Raman features observed at energies
close to the charge-transfer
gap [1]. To interpret the results obtained in [1] we suppose
that there exists
 an isolated bound state $|s>$ of a hole and a doubly occupied site
(doublon)
inside the Hubbard band.
The energy of this state is by $E_{b}\sim J$ lower than the optical
absorption threshold
$\omega_T\approx U$ corresponding to the location
of the upper Hubbard band. It follows from the Eq.(9)
that  at $\Omega\sim\omega_T-E_{b}$ the leading contribution to the
resonant cross section  comes from the $A_{2}$
scattering.
Thus we obtain that
the observable Raman intensity
is determined by the
 matrix element of the current-current commutator taken between the
ground state and the excited bound state
\begin{eqnarray}
R_{A_{2}}(\Omega)
\sim
|<0|[j_{x},j_{y}]|s>|^2\delta(\Omega-\omega_T+E_{b})
\label{R5}
\end{eqnarray}
Due to a semiclassical character of light scattering at large laser energy
the locations of the
currents $\vec j(\vec r)$ in the Eq.(16) have to
belong to the same diagonal of a
plaquette, otherwise in the leading order
in $\omega^{-1}_i$ the matrix element vanishes.

We characterize the bound state  by hole and
 doublon coordinates
and by a spin configuration  $|\{\sigma\}>$
which locally differs from the
ground state: $|s>=\int d\vec r d\vec r^\prime
\Psi(\vec r,\vec r^{\prime}) c^{\dag} (\vec r) c(\vec r^\prime)
|\{\sigma\}>$ where $\Psi(\vec r,\vec r^{\prime})$
 is the bound state wave function.
Then the expectation value of the current commutator acquires  the form
 \be
\la{R6}
|<0|[j_{x},j_{y}]|s>|^2
\approx 4|\int d{\vec r}
<0|T_{-x}T_{-y}T_{x+y}-T_{-y}T_{-x}T_{x+y}|0>\Psi(\vec r,\vec r+\hat
x+\hat y)|^2
\ee
The spin distorsion due to the
 presence of a localized bound state extends to very few plaquettes.
Therefore we can neglect
 a difference between the ground and the
excited spin configurations when calculating an overlap
factor $<0|\{\sigma\}>$. Similar to the case of an elastic scattering
(14,15), the expectation value
of the holonomy operator is equal to the solid angle subtended by three
spins belonging to the same plaquette
\be
\la{SSS}R_{A_2}(\Omega)\sim
\sum_{P}{\bigl(<0|(\vec S_1\times \vec S_2)\cdot \vec
S_3|0>\bigr)}^2\delta(\Omega-\omega_T+E_{b})
|\Psi(1,3)|^2
\ee
where $(1,2,3)$ are three neighbouring sites on a plaquette such as $1$ and $3$
belong to the same sublattice.
An essential difference, however, is that a resonant scattering
measures a spin chirality
locally- just in a position of a charged excitation while an
elastic scattering (Eq.(15)) gives the information about its spatial average.

Similarly,  the  other symmetries  contribute in the next
order in $\omega_i^{-1}$ and are independent of chirality:
\bea \la{R7}
R_{A_1}\sim
({{U-\Omega}\over {\omega_i}})^2\sum_{\mu=x,y}\int d\vec
r |<0|T_{-\mu}T_{-\mu}T_{2\mu}|\sigma>\Psi(\vec r,\vec r+2\vec {\mu})|^2,\\
R_{B_2}\sim
({{U-\Omega}\over {\omega_i}})^2\int d\vec
r |<0|T_{-x}T_{-y}T_{x+y}+T_{-y}T_{-x}T_{x+y}|0>\Psi(\vec r,r\vec r+\vec
x+\vec y)|^2\eea
These  can be also expressed in terms of spins according to the formula (10).

7.{\it Evidence of a new magnetic order}

Equations (17) and (18) lead to the following conclusion:
{\it the Raman data [1]
provide an evidence that the ground state is characterized
by long range correlations
between triads of adjacent
spins}. It is important to notice that this kind of order does not
 mean a parity
breaking. The simplest possibility to get a nonzero holonomy, while saving the
ground state invariance under the largest subgroup of the
magnetic class including parity, is presented  by the so-called
$\pi$-flux state.
In this state  $<0|\hat{\Phi}|0>=\pi$ on every plaquette
and translation operators  anticommute
 \bea \la{anti}
&&<0|T_xT_yT_{-x-y}|0>=-<0|T_{-y}T_xT_{-x+y}|0>=i\Delta,\nonumber\\
&&(T_xT_y+T_yT_x)|0>=0 \eea
In this state  the solid angle of three adjacent
spins {\it alternates} from one plaquette to another,
so its spatial  average staying in (15) vanishes
\bea \la{x}
&&<0|(\vec S_1\times \vec S_2)\cdot \vec S_3|0>=
-<0|(\vec S_{1^{\prime}}\times \vec
S_{2^{\prime}})\cdot\vec S_{3^{\prime}}|0>=\Delta,\nonumber\\
&&\sum_P <0|(\vec S_1\times \vec S_2)\cdot \vec S_3|0>=0\eea
Here $\{1,2,3\}$ and $\{1^{\prime},2^{\prime},3^{\prime}\}$ label any three
nearest sites on adjacent plaquettes with the same orientation. Let us note,
that the three-spin order (22) does not contradict with
 the antiferromagnet long
range order. Moreover it is conceivable that this kind of ordering
is realized in a $S=1/2$
Heisenberg model on a square lattice and can be revealed
 by means of the spin wave theory,
although  further analysis is necessary.

Note also that according to (20), it should
 be no scattering in the $B_2$ geometry if translations
do
anticommute  (see (21)). Indeed
no $B_2$ contribution to the
 resonant scattering at $\Omega\approx U$ has been detected in [1].

Next we argue that the  magnetic ground
state with the three-spin order (22) which is symmetric under
reflection: i) supports bound holon-doublon states ii)
 the bound state
by itself is odd under spatial reflection as well as time inversion.

8.{\it Zero mode}

The observation of the Raman resonance in the
$A_2$ geometry (16-17) leads to the
conclusion:  the equal time commutator $[j_{x},j_{y}]$ has a nonzero
matrix element between the ground state  and an excited state
$|s>$.
In turn, it
means that the ground state and the excited state $|s>$
have different parities. At the first glance it seems impossible because all
eigenstates of the Hubbard Hamiltonian on a square lattice are doubly
degenerate as a consequence
of the  reflection symmetry $\hat R$. Namely, if there is an
eigenstate  $|\Psi>$ with
currents $j_{x}|\Psi> ,<\Psi|j_{y}$, then
there is also an eigenstate $\hat R|\Psi>$  with currents
$j_{y}|\Psi>,<\Psi|j_{x}$. Thus
commutator $[j_{x},j_{y}]$ vanishes
 \be \la{jj} [j_x,j_y]=\sum_{|\Psi>}
\{j_{x}|\Psi><\Psi|j_{y}-j_{y}|\Psi><\Psi|j_{x}\}=0
\ee
An obvious way out
this problem is to assume a spontaneous parity
breaking in the
ground state: $\hat R|0>\ne |0>$.
 Once parity is broken a topological magnetic excitation $|\sigma>$
restores parity locally and then
it contributes to the $A_2$ scattering. This mechanism
is supposed to work in the case of the "chiral spin liquid"
state  proposed in [9-11].  Although this scenario seems to be unlikely,
particularly, because it is incompatible
 with the symmetry of the square lattice, and
also because of the lack of experimental evidence, we shall briefly discuss it
to contrast with our mechanism.

The chiral spin liquid state is
characterized by the uniform
expectation value of the chirality operator on each
plaquette
 \be \la{ttt}<0|T_yT_xT_{-x-y}|0>=<0|(\vec S(\vec r)
\times  \vec S(\vec r+\vec x)) \cdot \vec S(\vec r+\vec x+\vec y)|0>= \Delta
\exp(i{{\hat \Phi_{0}}\over 2}),\ee
where
\be\la{Delta}
|\Delta|^2=|<0|1+4\vec S_1\cdot\vec S_2+4\vec S_1\cdot\vec S_3+ 4\vec
S_3\cdot\vec S_2|0>|^2+|8<0|(\vec S_1 \times  \vec S_2) \cdot \vec S_3|0>|^2
\ee
The magnetic excitation $|\sigma>$ locally
reduces the value of the ground state
chirality by canting  spins at the location of
the
excitation.
As a result,
the sum in (23) doesn't vanish
which gives the $A_2$ Raman intensity
 \be
R_{A_{2}}(\Omega)\sim
\sum_{P}{\Delta}^2|<0|\sin{\hat \Phi_{0}\over 2}|0>|^2\delta(\Omega -U)
\ee

 Fortunately, it is not necessary to break parity in the ground state. It
may happen that there exists an excited state $|\Psi_0>$
annihilated
by the reflection operator: ${\hat R}|\Psi_0>=0$ which
has no reflection partner. This kind of state is known as a {\it zero
mode} [12]. Indeed a zero mode having no parity partner
could be the only
state contributing to the current commutator
\be \la{jj} [j_x,j_y]=
j_{x}|\Psi_0><\Psi_0|j_{y}-j_{y}|\Psi_0><\Psi_0|j_{x}\ee
Below we shall show that a zero mode is always a bound state
(with the energy of order
$U$) which appears inside the Hubbard gap.

In absence of an
experimental evidence in favor of any other parity
odd states we  identify the $A_2$ final  state $|s>$
as the zero mode with energy
$E_b=\omega_T -U \sim J$. In fact it is a bound
state of
a hole and a doublon in presence of a topological spin soliton. As we
have seen the existence of the zero mode resulting to a nonvanishing
commutator $[j_x,j_y]$  dictates unambiguously the main  property of the
magnetic ground state:
 holonomies (21) must be nontrivial.
A more complicated analysis [7]
shows that this is also sufficient for
the described magnetic ground state
 to possess a zero mode.

Eventually, the zero mode contribution which
 appears to be the only one present in the case
of the resonant ${A_{2}}$
 scattering (16-18) can be estimated as
\be
\la{R7}
R_{A_{2}}(\Omega)
\sim 4\sum_{\vec r}
\Delta^2|\Psi_0(\vec r)|^4\delta(\Omega-U)
\ee
where $\Delta$ is given by (25).

9. {\it Adiabatic approximation}

In order to see that a zero mode is always a bound state
( in fact the lowest
bound state) let us consider a charged  excitation
in a slowly varying (adiabatic) spin background. Following the
Ref. [13-14] we describe  a charged excitation in the Mott
insulator by a coherent state of a hole
$\psi(\vec r)$ and a hard core boson $z_{\sigma}(\vec r)$
representing a local
spin:
$<0|\vec S(\vec r)|0>=\bar z_{\sigma}(\vec
r) \vec\sigma_{\sigma \sigma^{\prime}}z_{\sigma^{\prime}}(\vec r)$
with the constraint $|z_1|^2+|z_2|^2=1-\psi^{\dag}\psi$.

In terms of these variables the translation operator can be written as
$T_{\mu}(\vec r)=\bar z(\vec r+\hat \mu)z(\vec r)$
and the $t-J$ Hamiltonian now reads
 \be
\la{ham1}
H=\sum_{\vec a,\vec b}t\Delta_{\vec a,\vec b}\psi^{\dag}(\vec a)\psi(\vec
b)+c.c. +H_J
\ee
where  $$H_J=\sum _{\vec r}U\psi^{\dag}(\vec r)\psi(\vec r)+\sum _{\vec a,\vec
b}
J(1-\psi^{\dag}(\vec
r)\psi(\vec r)) \vec S_{\vec a} \vec S_{\vec b}$$
are Coulomb and magnetic terms in the Hamiltonian,
$\Delta_{\vec a,\vec b}=\bar z_{\sigma}(\vec a)z_{\sigma}(\vec b)$ and
$\vec a$ and $\vec b$
 denote nearest neighbouring sites on the
$A$ and $B$ sublattices of the square
lattice. Since in the ground state  spins
are approximately
antiparallel the energy of a holon-doublon pair located on the nearest sites
is by order of $J$ higher than if they were put on
the next nearest sites.  If the hopping energy $t$ is less than the exchange
energy
 $J$  a hole and a doublon appear on different sublattices only
virtually. Therefore in the leading  order in $t/J$ one may consider only
processes of two
subsequent jumps described by the effective Hamiltonian
\be
\la{ham2}
H=\sum_{\vec a,\vec a^{\prime},\vec b,\vec b^{\prime} }
t^{\prime}\{\psi^{\dag}(\vec a)\Delta_{\vec a,\vec b^{\prime}}
\Delta_{\vec b^{\prime},\vec a^{\prime}}
\psi( \vec a^{\prime}) +\psi^{\dag}(\vec b)\Delta_{\vec b,\vec a^{\prime}}
\Delta_{\vec a^{\prime},\vec b^{\prime}}
\psi( \vec b^{\prime})\}+H_J \ee
where $t^{\prime}=-{t^2}/\sum_{\mu} J<0|\vec S(\vec r)\cdot\vec S(\vec r+\hat
{\mu})|0>$. Now the  effective two-step hopping processes
 within one sublattice
only slightly
change a spin configuration and therefore  these can be treated
adiabatically. Namely, we may consider
 the Schrodinger
equation  for  a hole  and a doublon
in an equilibrium
spin configuration.
The Schrodinger operator in the Eq.(30) is a square of the
elementary translation operator, so we may apply a standard argument
 about a zero mode [15].
 Since the effective hopping operator (30) is a
hermitian operator squared,
the energy of any state is positive or zero. All states with
nonzero energy are doubly degenerate. Let us write down the eigenstates in the
sublattice basis. If $|\Psi>=(\Psi(\vec a),\Psi(\vec b))$ is an
eigenstate  then its
partner $|\tilde\Psi>=(\sum_{\vec b}\Delta_{\vec a,\vec b}\Psi(\vec
b),-\sum_{\vec a}\Delta_{\vec b,\vec a}\Psi(\vec a))$
which is obtained by the
time  reversal transformation is also an eigenstate
with
the same energy.  Since in the half-filled case
each of these states
remains
 invariant under the charge conjugation we conclude that
these can be transformed to each other by parity transformation
( spatial reflection) and then their contributions
to the current-current commutator (23) cancel out. The only nondegenerate
state could be that one which has a zero energy and is
annihilated by one of the operators
$\Delta_{\vec a,\vec b}$ or $\tilde\Delta_{\vec
b,\vec a}$. Since $ <0|(\vec S_1\times \vec S_2)\cdot \vec S_3|0>\ne 0$
the operators
$\Delta_{\vec a,\vec b}$ and $\tilde\Delta_{\vec
b,\vec a}$ do not commute, so only one of them may have a zero mode
for a given spin configuration. Thus there are
two candidates: (i) a zero mode $\Psi_0(\vec b)$  of the operator
$\Delta_{\vec a,\vec b}\Psi_0(\vec b)=0$ and $\Psi_0(\vec a)=0$ and
(ii) a zero mode $\tilde\Psi_0(\vec a)$ of the operator
 $\tilde\Delta_{\vec b,\vec a}\tilde\Psi_0(\vec a)=0$ and
$\tilde  \Psi_0(\vec b)=0$. In each of the two cases the zero mode
is nondegenerate and breaks the two dimensional
parity.
Note that
charge excitations occupying
the zero mode state stay on only one of the two sublattices.

All other states including possible bound states have higher
energies. Therefore we conclude that (i) if  a zero mode   exists then
it appears to be the
midgap state, (ii)  it exists only
 if translations  in the ground state do not
commute. Since the magnetic ground state is assumed to be parity
even the translations
must anticommute (21).

Many properties of zero modes are universal and depend only on the
crystal symmetry of the lattice. They
constitute the  subject of the "index" theorem.
In particular, the index
theorem states
 that the spin
configuration which supports
a zero mode must have a nontrivial topology.

This gives us to a simple geometrical interpretation of the
 spin configuration
surrounding a bound state of a hole and a doublon. It is a magnetic
hedgehog which carries one quantum of the flux on the top of the ground state.
One quantum
of the topological charge creates a zero mode which
can accomodate one hole and one doublon. Moreover a local
"holon" density of the occupied zero mode is
given by the topological charge density  of the magnetic hedgehog
\be \la{tc}
 q(\vec r)={1\over{2\pi}}[ {d\over{dx}} \vec n(\vec r) \times
{d\over{dy}} \vec n(\vec r) ] \cdot \vec n(\vec r)
\ee
where $\vec n(\vec r)$ describes a spin configuration on the sublattice
which carries charge excitations.
Finally, combining Eqs.(17) and (32) we obtain
that the current-current commutator
possesses the anomalous Schwinger
term  given by a topological density of the spin configuration
\be
\la{R8}[j_x(\vec r),j_y(\vec r^{\prime})]|s>\sim iq(\vec r)
\delta(\vec r - \vec r^{\prime})|s> \ee
which  appears explicitly in the scattering rate (28).

10.{\it Zero mode and topological mechanism of superconductivity}

A presence of zero modes
in the insulating state implies remarkable properties
of the doped state. Due to the particle-hole symmetry of the half filled band
 the spin configuration $|s>$ in presence of two holes
contains a spin soliton
 in the same way as in the case of a holon-doublon pair. Therefore at
small doping which does not destroy the magnetic ground state
two holes couple with each other inside the topological spin bag.
A residual
interaction between zero modes  lifts their degeneracy and
opens a narrow midgap band which is a cornerstone of the
topological mechanism of superconductivity. Although the
midgap band is completely filled it always remains
compressible. To show this let us suppose that we insert
two additional holes. To get accomodated to the midgap band
these will create a topological soliton increasing a midgap band
capacity
by an extra zero mode state. The corresponding  energy cost
is proportional to doping and the midgap band remains completely
filled.

The most significant feature of the topological superfluid state is its
current algebra including  the
anomalous term similar to Eq.(32) which is  now realized
in the (doped) ground state
$$<0|[j_x(\vec r),j_y(\vec r^{\prime})]|0>
\sim <0|\rho(\vec r)\delta(\vec r - \vec r^{\prime})|0>
\sim q(\vec r)\delta(\vec r - \vec r^{\prime})$$

11.{\it Elastic Raman scattering in the Quantum Hall effect}

As another comment we also state
that an anomalous current algebra gives rise to a nontrivial
elastic light scattering
in the Quantum Hall Effect.
 For simplicity we consider the case of the Integer  Quantum Hall Effect
of spin polarized fermions which allows us to
ignore  any intra-Landau level excitations as well as
a magnetoroton mode.
 Then the spectrum contains only charge transfer
inter-Landau level
excitons with a gap equal to the cyclotron frequency
$\omega_c$. According to our previous discussion
the  light scattering should be observed at zero energy transfer,
i.e. inside the gap where there are no real states. This happens due to the
anomalous equal time current-current  commutator
in an external magnetic field
$$[j_x,j_y]= i{{2\mu}\over m}B$$
where $\mu$ is the Bohr magneton, $m$ is an electron mass and B is the
magnetic field strength.
Substituting this expression to the Eq.(7) we get
$$R(\Omega)= {{{\omega_s}^3\omega_i}\over {c^4}}B^2\delta(\Omega)$$
Thus a nontrivial  elastic Raman scattering can be found even in
an incompressible liquid provided its  current algebra is anomalous.
In fact,
this phenomenon is quite similar to the
dissipationless Hall current in an
incompressible magnetized electron liquid.

12.{\it Experiment}

Below we discuss some features of the experiments [1] which  motivated
the proposed mechanism.

 The Raman measurements  were made on insulating cuprates $
Y(Pr)Ba_2Cu_3O_{6+x}$ and $Gd(Nd)_2CuO_4$. A strong $A_2$ peak has been
observed in all these
materials about $0.2 ev$ below the optical absorption peak. This
value which is of the order of the
magnetic exchange energy $J$
gives the distance between the zero mode energy level
and a continuous part of the spectrum (upper Hubbard band).
We consider it as a strong
 indication that the  observed phenomenon has a  magnetic
origin and is due to the many-body effects.
It should be contrasted to any mechanism based on local
interatomic transitions between oxygen and copper orbitals.

Beyond the adiabatic approximation the $A_2$  peak broadens
and acquires a width of the
order of the
inverse spin relaxation time which  remains of the order of $J$
 even at low
temperatures. This does not contradict with the  experiments [1] which
show that the width of the
peak
remains of the same order and only
slightly narrows as temperature decreases.

A
weaker Raman
feature was resolved in the $A_1$ geometry
at approximately  the same energy as the $A_2$ peak,
while no $B_2$ feature
was found. According to the eq.(19) the $A_1$ scattering
rate is indeed less then the $A_2$ rate by the factor $({J\over \omega_i})^2$.
 The vanishing of the
$B_2$ rate (20) is another strong indication in favor
of the anticommutativity of
translations (see (21)).

 As we saw the anomalous $A_2$ peak is quite universal.
It primarily depends on  the symmetry of the magnetic structure  and does
not depend on details of the Hubbard-type
Hamiltonian. On the contrary,
 the $A_1$ scattering rate (19) does depend on details of the Hamiltonian.
 These features were also observed and especially stressed in
the Ref.[1] - the
$A_1$ intensity changes among  different substances
and, in particular, it was not observed in $YBa_2Cu_3O_{6.1}$.

13. {\it Summary}

In summary, we showed that inelastic light scattering in the Mott insulator
at large energy transfers provides
 a direct way to measure the current algebra and
the symmetry of charged excitations. The observation of Raman features
in insulated cuprates
below the optical absorption threshold in cross polarizations is a
strong indication on the existence
of an anomalous current algebra and midgap
states (zero modes)
formed by topological spin solitons bounded with two charged excitations
(a hole and a doubly occupied site). This observation
suggests that in the doped case electronic currents also
obey the anomalous algebra and two holes also bind together with
 a topological soliton.

{\it Acknoledgements}

 We are indebted to M.Klein who informed us about
the data of his group prior to publication.
We also acknowledge valuable discussions with A.V.Chubukov,
D.Frenkel,  B.Shraiman, P.Sulewski and A.Pinczuk. This work was
supported by the Science and Technology
Center for Superconductivity through the Grant NSF-STC-9120000.

\pagebreak


\begin{references}
\bibitem{Klein} Ran Liu, D.Salamon, M.V.Klein, S.L.Cooper, W.C.Lee,S.W.Cheong,
and D.M.Ginsberg, Phys.Rev.Lett.{\bf 71}, 3709 (1993).
\bibitem{rev} S.L.Cooper and M.V.Klein, Comments on Condensed Matter Physics,
{\bf 2}, 99 (1990).
\bibitem{exp} K.Lyons, P.Sulewski, P.Fleury, H.Carter, A.S. Cooper,
G.Espinosa, Z.Fisk, and S.Cheong,
 Phys.Rev. {\bf 39}, 9693 (1989).
\bibitem{SS} B.S.Shastry and B.I.Shraiman,
 Phys.Rev.Lett. {\bf 65}, 1068 (1990); Int.J.Mod.Phys.{\bf B5}, 365 (1991).
\bibitem{CF} A.V.Chubukov and D. Frenkel, to appear in Phys.Rev.B.
\bibitem{L} R.Laughlin,Phys.Rev.Lett., {\bf 60},2677 (1988).
\bibitem{W1} P.B.Wiegmann, Progress of Theoretical Physics,
Supplement {\bf 107}, 243 (1992); Princeton preprint IASSNS-HEP-91/29.
\bibitem{LL} L.D.Landau and E.M.Lifshitz, {\it Course of Theoretical Physics,
v.7} ((Pergamon Press, 1968).
\bibitem{WWZ} X.G.Wen, F.Wilczek and A.Zee, Phys. Rev. B {\bf 39}, 11413
(1989).
\bibitem{KW} D.V.Khveshchenko and P.B.Wiegmann, Mod.Phys.Lett. B {\bf 3}, 1383
(1989); ibid {\bf 4}, 17 (1990).
\bibitem{W2} P.B.Wiegmann, Nobel Symp.73, Phys. Scr.,{\bf T27}, (1988).
\bibitem{zm} Zero modes  appeared in condensed matter theory in the context of
Charge Density Waves and Peierls instability and are sometimes
 called  midgap states. See, e.g.
S.P.Brazovsky and N.Kirova, Sov.Sci.Rev. Harwood Acad.Publ.{\bf A5}, 99 (1984);
A.J.Heeger, S.Kivelson and W.-P.Su, Rev.Mod.Phys.{\bf
60}, 781 (1988). Its relevance for doped Mott insulators was pointed out
 in [7].
\bibitem{SiS} B.I.Shraiman and E.D.Siggia,
 Phys.Rev.Lett. {\bf 61}, 467 (1988).
\bibitem{W3} P.B.Wiegmann, Phys.Rev.Lett. {\bf 60}, 821 (1988).
\bibitem{Wit} E.Witten, Nucl.Phys. {\bf B185}, 513 (1981).
\end{references}
\end{document}